\documentclass{PoS}

\pdfoutput=1

\usepackage{graphicx}
\usepackage{amsmath}
\usepackage{amssymb}
\usepackage{float}
\usepackage{appendix}
\usepackage{listings}
\usepackage{subcaption}
\usepackage{cite}

\newcommand{\stringa}{\ttfamily\lstinline}
\def\cod#1{{\stringa!#1!}}

\title{Azimuthal-angle Observables in Inclusive Three-jet Production}

\ShortTitle{Azimuthal-angle Observables in Inclusive Three-jet Production}

\author{\speaker{G. Chachamis}\\
        Instituto de F{\' \i}sica Te{\' o}rica UAM/CSIC, Nicol{\'a}s Cabrera 15\\
        \& Universidad Aut{\' o}noma de Madrid, E-28049 Madrid, Spain.
        E-mail: \email{chachamis@gmail.com}}

\author{F. Caporale\\
        Instituto de F{\' \i}sica Te{\' o}rica UAM/CSIC, Nicol{\'a}s Cabrera 15\\
        \& Universidad Aut{\' o}noma de Madrid, E-28049 Madrid, Spain.
        E-mail: \email{francesco.caporale@uam.es}}
        
\author{F.~G. Celiberto\\
        Dipartimento di Fisica, Universit{\`a} della Calabria \&\\
        Istituto Nazionale di Fisica Nucleare, Gruppo Collegato di Cosenza,\\
        I-87036 Arcavacata di Rende, Cosenza, Italy.
        E-mail: \email{francescogiovanni.celiberto@fis.unical.it}}

\author{D. Gordo G{\' o}mez\thanks{La Caixa-Severo Ochoa Scholar.}\\
 Instituto de F{\' \i}sica Te{\' o}rica UAM/CSIC, Nicol{\'a}s Cabrera 15\\
        \& Universidad Aut{\' o}noma de Madrid, E-28049 Madrid, Spain.
        E-mail: \email{david.gordo@csic.es}}

\author{A. Sabio Vera\\
 Instituto de F{\' \i}sica Te{\' o}rica UAM/CSIC, Nicol{\'a}s Cabrera 15\\
        \& Universidad Aut{\' o}noma de Madrid, E-28049 Madrid, Spain.
        E-mail: \email{a.sabio.vera@gmail.com}}

\abstract{We discuss the impact of corrections beyond the leading-logarithmic accuracy on some recently 
proposed LHC observables that are based on azimuthal-angle ratios in a kinematical setup that is an extension to
 the usual one for Mueller-Navelet jets, after requiring an extra tagged jet in central regions of rapidity. 
 The corrections tend to be mild which suggests that these observables are an excellent way to probe the onset of BFKL effects
 at hadronic colliders.}

\FullConference{ 25th International Workshop on Deep Inelastic Scattering and Related Topics \\
                 3-7 April 2017 \\
                 University of Birmingham, Birmingham, UK}

\begin{document}

\section{Introduction}

In high energy scattering 
 in Quantum Chromodynamics (QCD) 
 the Balitsky-Fadin-Kuraev-Lipatov (BFKL) framework in the leading logarithmic (LLA)~\cite{Lipatov:1985uk,Balitsky:1978ic,Kuraev:1977fs,Kuraev:1976ge,Lipatov:1976zz,Fadin:1975cb} and next-to-leading logarithmic (NLLA) approximation~\cite{Fadin:1998py,Ciafaloni:1998gs} are used for the resummation of large logarithms in the center-of-mass energy squared $s$ 

An interesting case is Mueller-Navelet jets~\cite{Mueller:1986ey}, configurations
with two final state jets\footnote{
 Another interesting idea,  suggested in~\cite{Ivanov:2012iv} 
and investigated in~\cite{Celiberto:2016hae,Celiberto:2017ptm}, is
the study of the production of two charged light hadrons,
$\pi^{\pm}$, $K^{\pm}$, $p$, $\bar p$, with large transverse momenta and well
separated in rapidity.} with transverse momenta of similar sizes, $k_{A,B}$,  
and a rapidity distance $Y=\ln ( x_1 x_2 s/(k_A k_B))$ large.
A number of studies~\cite{DelDuca:1993mn,Stirling:1994he,Orr:1997im,Kwiecinski:2001nh,Andersen:2006pg,DeRoeck:2009id,Angioni:2011wj,Caporale:2013uva, Caporale:2013sc,Marquet:2007xx,Colferai:2010wu,Ducloue:2013wmi,Ducloue:2014koa,
Mueller:2015ael,Chachamis:2015crx,N.Cartiglia:2015gve}   
focused on the azimuthal angle ($\theta$) behaviour of the two jets
regards that the presence of decisive minijet activity in the rapidity space between these two outermost jets  can
be accounted for by a BFKL gluon Green function connecting the two jets.
Moreover, it was shown~\cite{Vera:2006un,Vera:2007kn}, 
that ratios of projections on azimuthal angle observables
\begin{eqnarray}
{\cal R}^m_n = \langle \cos{(m \, \theta)} \rangle / \langle \cos{(n \, \theta)} \rangle\,\,,
\end{eqnarray}
(where $m,n$ are integers) are much more favourable quantities in the search for a clear signal of BFKL effects. The comparison of different NLLA
calculations for these ratios ${\cal R}^m_n$~\cite{Ducloue:2013bva,Caporale:2014gpa,Caporale:2014blm,Celiberto:2015dgl,Celiberto:2016ygs} against LHC experimental data has been promising.

Along these lines, new observables were recently proposed for processes with three-jet~\cite{Caporale:2015vya,Caporale:2016soq}
and four-jet final states~\cite{Caporale:2015int,Caporale:2016xku} with the outermost jets having a large rapidity distance
and any other tagged jet is to be found in more central regions of the detector. Here we will restrict the discussion to inclusive 
three-jet production assuming that the jets are connected in the $t$-channel via gluon Green's functions.
The main idea presented in Refs.~\cite{Caporale:2015vya,Caporale:2016soq} was to produce theoretical estimates for the 
ratios 
\begin{eqnarray}
R^{M N}_{P Q} =\frac{ \langle \cos{(M \, \theta_1)} \cos{(N \, \theta_2)} \rangle}{\langle \cos{(P \, \theta_1)} \cos{(Q \, \theta_2)} \rangle} \, = \, \frac{C_{MN}}{C_{PQ}} \,, 
\label{Rmnpq}
\end{eqnarray}
where $\theta_1$ is the azimuthal angle difference between the first and the second (central) jet, while,
$\theta_2$ is the azimuthal angle difference between the second  and the third jet.
In this work, we show theoretical predictions for $R^{22}_{33}$ at NLLA, originally presented in~\cite{Caporale:2016zkc}.
We are interested in seeing whether we have large corrections once we consider the NLLA gluon Green's functions 
instead of the leading-logarithmic accuracy ones. Large corrections of that origin is not a surprising outcome for many BFKL-based calculations and they could potentially have a strong impact on the ratios $R^{M N}_{P Q}$, hence they need to be addressed.

We assume for the outermost jet transverse momenta that
$k_A^{\rm min} = 35$ GeV, $k_B^{\rm min} = 50$ GeV,  $k_A^{\rm max} = k_B^{\rm max}  = 60$ GeV,
whereas  the transverse momentum of the central jet, $k_J$, can live in three wide bins, that is, ~$20\, \mathrm{GeV} < k_J < 35\, \mathrm{GeV}$ (bin-1),
$35 \,\mathrm{GeV} < k_J < 60\, \mathrm{GeV}$ (bin-2) and
$60\, \mathrm{GeV} < k_J < 120\, \mathrm{GeV}$ (bin-3). 
To quantify the difference between LLA and NLLA, we define 
\begin{eqnarray}
\delta x(\%) = \left(
 \text{res}^{\rm(LLA)} - \frac{\text{res}^{\rm (BLM-1)}+\text{res}^{\rm (BLM-2)}}{2}
 \right) \frac{1}{ \text{res}^{\rm(LLA)}}\,\,,
 \label{corrections}
\end{eqnarray}
where  $\text{res}^{\rm(LLA)}$ is the LLA result while $\text{res}^{\rm(BLM-1)}$ and $\text{res}^{\rm(BLM-2)}$
are NLLA results in the BLM scheme~\cite{BLM} that differ in the renormalisation scale to give us a measure of the 
theoretical uncertainty.

\section{Results}
\begin{figure}[H]
\hspace{-.9cm}
\begin{subfigure}{.1\textwidth}
\centering
   \includegraphics[scale=0.29]{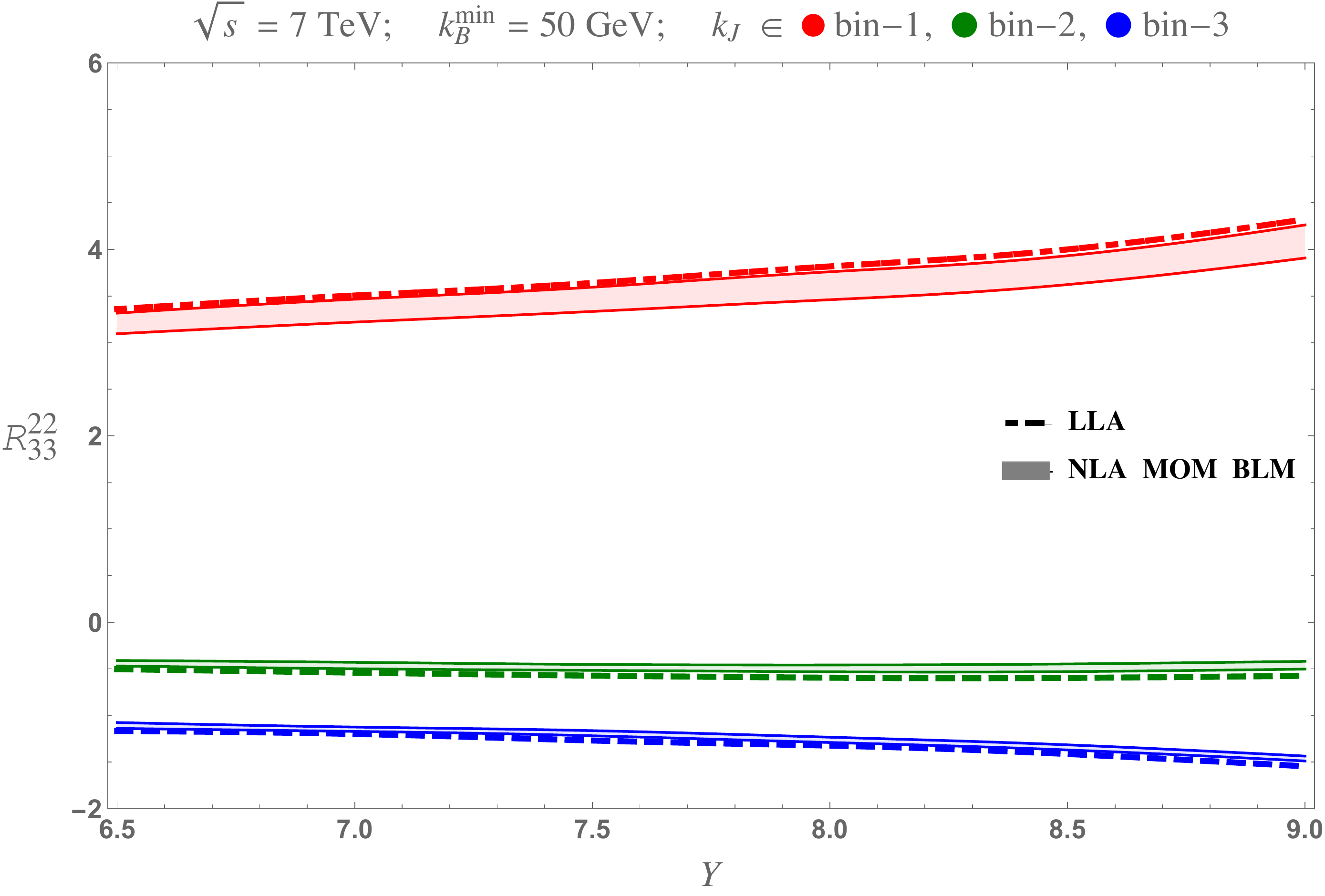}
\end{subfigure}
\begin{subfigure}{1.42\textwidth}
\centering
   \includegraphics[scale=0.29]{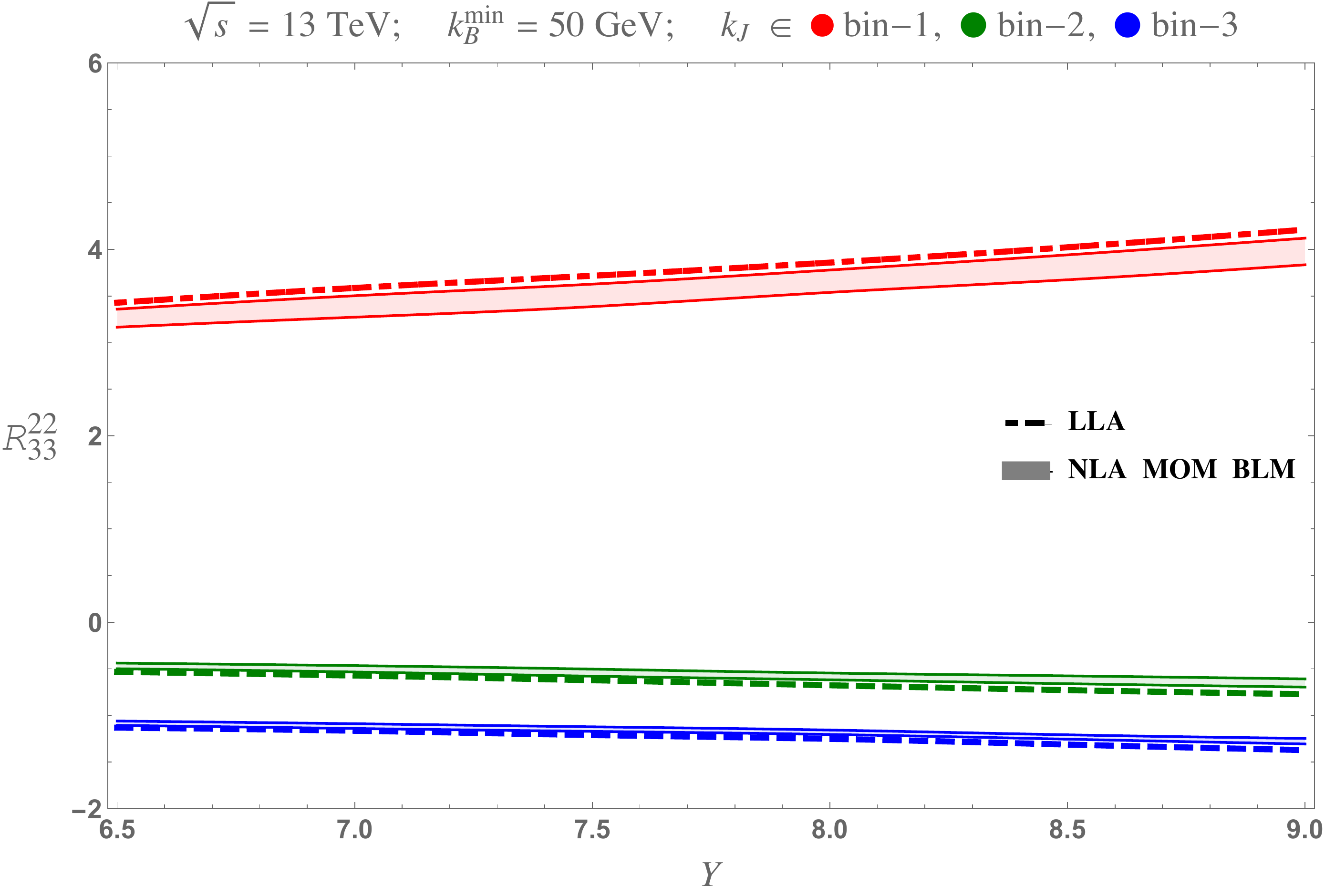}
\end{subfigure}
\caption{\small $Y$-dependence of the LLA (dashed lines) and NLLA (bands) results for 
$R^{22}_{33}$ with $y_J$ integrated over a central rapidity bin at $\sqrt s = 7$ TeV
 (left) and $\sqrt s = 13$ TeV (right).} 
\label{fig:first}
\end{figure}

First we present results for $R^{22}_{33}$ as a function of the rapidity difference after integrating over a central jet rapidity bin, that is,
after allowing for the central jet rapidity to take values in the range  $-0.5 < y_J < 0.5$, see Fig.~\ref{fig:first}.
We use dashed lines to represent LLA results and a band for the NLLA results. The band is bounded by the
$\text{res}^{\rm(BLM-1)}$ and $\text{res}^{\rm(BLM-2)}$ results in thin continuous lines.
The red curve and band are representing the LLA and NLLA results respectively for the central jet belonging to bin-1
the green curve and band for the central jet belonging to bin-2 and
the blue curve and band are representing the LLA and NLLA results respectively for the central jet belonging to bin-3. We see that the
NLLA corrections are mild. Moreover, we note that the overall picture does not change considerably when we go from LLA to NLLA.

\begin{figure}[H]
\hspace{-.9cm}
\begin{subfigure}{.1\textwidth}
\centering
   \includegraphics[scale=0.29]{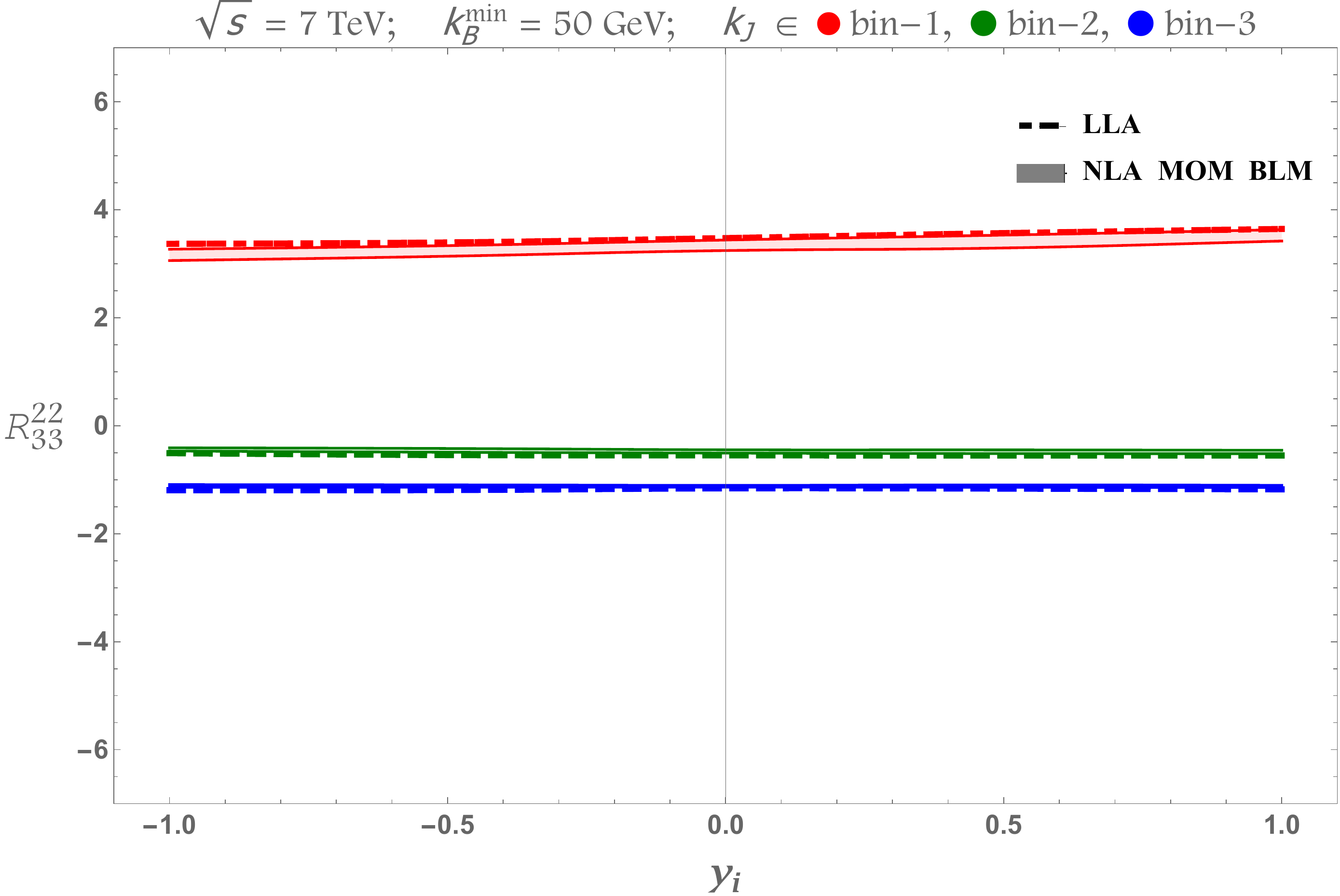}
\end{subfigure}
\begin{subfigure}{1.42\textwidth}
\centering
   \includegraphics[scale=0.29]{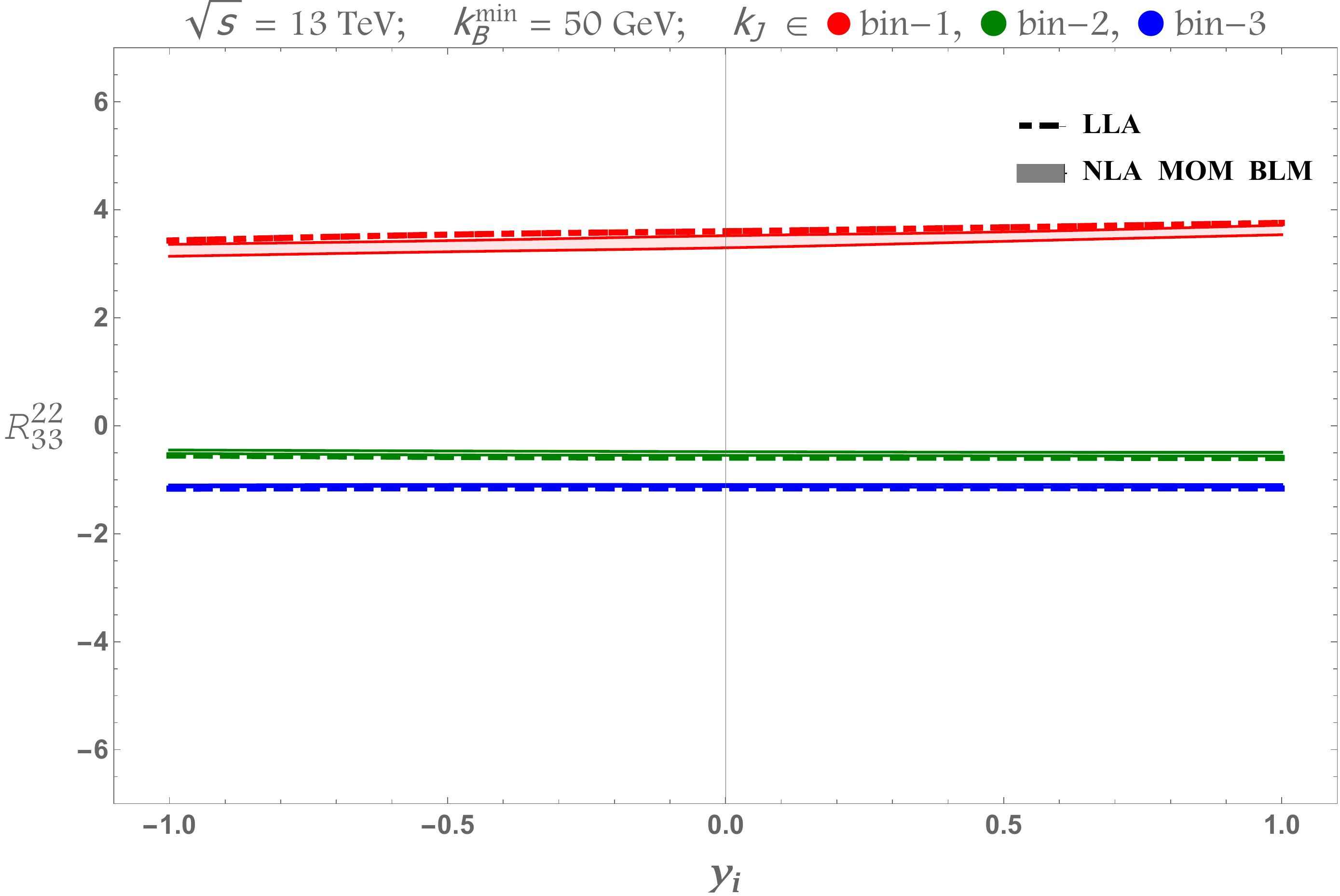}
\end{subfigure}
\caption{\small $y_i$-dependence of the LLA (dashed lines) and NLLA (bands) results for 
$R^{22}_{33}$  at $\sqrt s = 7$ TeV
 (left) and $\sqrt s = 13$ TeV (right).} 
\label{fig:second}
\end{figure}

Next,  in Fig.~\ref{fig:second}, we allow for $Y_A$ (rapidity of the forward jet) and $Y_B$ (rapidity of the backward jet)
to take
values such that $(Y_A^{\text{min}} = 3) < Y_A < (Y_A^{\text{max}} = 4.7)$ and
$(Y_B^{\text{min}} = -4.7) < Y_B < (Y_B^{\text{max}} = -3)$. Furthermore,
the rapidity of the central jet can take values in five distinct rapidity bins of unit width, that is,
$y_i-0.5 < y_J<y_i+0.5$, with $y_i = \{-1, -0.5, 0, 0.5, 1\}$.
Obviously,  the ratios now are functions of $y_i$, namely,
 \begin{eqnarray}
 R_{PQ}^{MN}(y_i) \, =\frac{ C_{MN}^{\text{integ}}(y_i)}{C_{PQ}^{\text{integ}}(y_i)}\, .
\end{eqnarray}
We see that the $y_i$-dependence of the three ratios is very weak.
Moreover, the similarity between the $\sqrt s = 7$ TeV and $\sqrt s = 13$ TeV
plots is more striking that in the Fig.~\ref{fig:first} and the NLLA corrections seem even milder.

\section{Summary \& Outlook}

We have shown some results from a first beyond the leading logarithmic accuracy work 
on generalised azimuthal-angle observables in
inclusive three-jet production at the LHC within the BFKL formalism. 
In addition, for a proper study of the total corrections beyond the LLA, the NLO jet vertices
need to be included
and the NLLA gluon Green functions. 

Our most important conclusion in this work is that the corrections
that come into play after considering NLLA gluon Green's functions are 
generally mild and the generalised ratio observables
exhibit perturbative stability. Moreover, we note that the differences 
in the plots for 7 TeV to 13 TeV are
small which suggests that
these observables capture the crucial features
of the BFKL dynamics with regard to
the azimuthal behavior of the hard jets in hadronic inclusive three-jet production.
We plan to compare our results here against theoretical estimates for these observables 
from fixed order calculations, the full BFKL Monte Carlo \cod {BFKLex}~\cite{Chachamis:2011rw,Chachamis:2011nz,Chachamis:2012fk,
Chachamis:2012qw,Caporale:2013bva,Chachamis:2015zzp,Chachamis:2015ico,Chachamis:2016ejm} as well as from
general-purpose Monte Carlos tools.

\begin{flushleft}
{\bf \large Acknowledgements}
\end{flushleft}
This work was supported by the Spanish Research Agency (Agencia Estatal de Investigación) through the grant IFT Centro de Excelencia Severo Ochoa SEV-2016-0597.
GC and ASV acknowledge support from the MICINN, Spain, under contract FPA2016-78022-P. 
DGG is supported with a fellowship of the international programme ``La Caixa-Severo Ochoa''.

\end{document}